# Spectral Extension and Synchronisation of Microcombs in a Single Microresonator

Shuangyou Zhang[1,2], Jonathan Silver[1,3], and Pascal Del'Haye[1,2*]

[1]National Physical Laboratory (NPL), Teddington, TW11 0LW, United Kingdom

[2]Max Planck Institute for the Science of Light, 91058 Erlangen, Germany

[3]City, University of London, London, EC1V 0HB, United Kingdom

*Corresponding author: pascal.delhaye@mpl.mpg.de

## Abstract

We demonstrate that the optical spectrum of a soliton microcomb generated in a microresonator can be coherently extended by using bichromatic pumping: one pump with a frequency in the anomalous dispersion regime of the microresonator is used to generate a bright soliton microcomb while a second pump in the normal dispersion regime both compensates the thermal effect of the microresonator and generates a second frequency comb via cross-phase modulation with the bright soliton, thus extending the spectrum of the microcomb. Experimental results for different microresonators show that the mode spacings of the two combs are synchronised, even when the combs are generated in different mode families. Numerical simulations agree well with experimental results and reveal that a bright optical pulse from the second pump is passively formed in the normal dispersion regime and trapped by the bright soliton. We believe the demonstrated technique provides an alternative way to generate broadband microcombs and multi-coloured frequency combs that can be useful for frequency metrology and spectroscopy. In addition, this method enables the selective enhancement of available optical power in specific parts of the comb spectrum.

## Introduction

Microresonator-based frequency combs ("microcombs") provide a promising platform for miniaturising optical frequency comb systems and have drawn significant attention in the past decade due to their compactness, high repetition rates from 10 GHz to 1 THz, and broad spectral bandwidths[1,2]. In particular, soliton formation in microresonators has been demonstrated to generate low noise, fully coherent frequency combs[3-5]. To date, microcombs have been successfully used for optical frequency synthesis[6], optical coherent communications[7], laser-based light detection and ranging[8,9], dual-comb spectroscopy[10,11], and low noise electronic signal generation[12,13], just to name a few.

Broadband optical frequency combs are required for precise optical frequency metrology and spectroscopy. Soliton microcombs have been demonstrated with spectral bandwidths reaching more than one octave[14-19], and octave-spanning soliton microcombs with THz mode spacing directly realised using $Si_3N_4$ resonators by carefully engineering their dispersion properties[14,15]. Spectra of soliton microcombs with GHz mode spacing can be externally broadened to cover an octave using highly nonlinear fibers[16] or nonlinear waveguides[18]. Bichromatic pumping of microresonators at similar wavelengths has been studied for microcomb generation with the benefits of a thresholdless pump intensity and stabilisation of the frequency comb repetition rate[20-24]. More recently, bichromatic pumping has been demonstrated for the generation of dual orthogonally polarised microcombs both theoretically and experimentally[25,26].

Here, we demonstrate the coherent extension of the spectral bandwidth of a soliton microcomb by bichromatic pumping. One of the pump lasers, within the C-band (1530 - 1565 nm), is used to generate a primary bright soliton microcomb, where the overall group velocity dispersion of the microresonator is anomalous. Another pump laser in the O-band (1260 - 1360 nm) is used both to compensate the thermal effect of the microresonator during soliton generation[27] and to generate a second frequency comb via Kerr cross-phase modulation (XPM) with the primary soliton pulse, which leads to an extension of the comb spectrum. Importantly, our results show that the comb line spacing of the second frequency comb has exactly the same value as that of the primary soliton microcomb. This has been found to be the case even when using differently-polarised optical modes for the two pumps, and for several different microresonators. Numerical simulations agree well with the experimental results and show that a bright optical pulse is passively formed and synchronised with the primary soliton pulse via XPM. We believe that our findings could find applications in multi-coloured frequency comb generation and optical frequency metrology.

## Results

Figure 1 shows the schematic of the experimental setup. A 1.5 μm wavelength external cavity diode laser (ECDL) with a short-term linewidth of <10 kHz is used as the first pump laser for generating a soliton microcomb while a similar ECDL at 1.3 μm is used as a second pump laser. Here, the second pump laser plays two roles: firstly to passively stabilise the circulating optical power inside the microresonator to facilitate soliton generation from the first pump[27], and secondly to generate a second frequency comb to extend the overall comb spectrum. For clarity, throughout this manuscript, the first pump at 1.5 µm is referred as to the primary pump and the second pump laser at 1.3 µm as the auxiliary pump. The inset in Fig. 1 shows the 250-µm-diameter fused silica microtoroid used in the experiments, with a free spectral range (FSR) of 257 GHz[28]. The microtoroid was fabricated from a silicon wafer with a 6-µm layer of silicon dioxide ($SiO_2$)[13]. A mode family of the resonator with a quality factor of $\sim 2 \times 10^8$ was chosen for the soliton generation, and the 1.5-µm primary pump laser was tuned to a wavelength around 1547.9 nm to generate the soliton microcomb. The 1.3-µm auxiliary laser is used to pump a mode of the same mode family at a wavelength of 1335 nm. The two lasers were combined with a wavelength division multiplexer (WDM) and evanescently coupled into the microresonator via a tapered optical fibre. Fibre polarisation controllers (PCs) were used to match the polarisations of each laser to the modes of the microresonator. At the resonator output, part of the light was monitored with an optical spectrum analyser (OSA). Another WDM was used to separate the two pump wavelengths in order to monitor their transmissions via two photodiodes (PD1 and PD2). In addition, a reflective diffraction grating was used to spatially select part of the optical spectrum of the microcomb and to subsequently detect the filtered optical spectrum with an ultrafast photodiode (PD3, 50 GHz bandwidth).

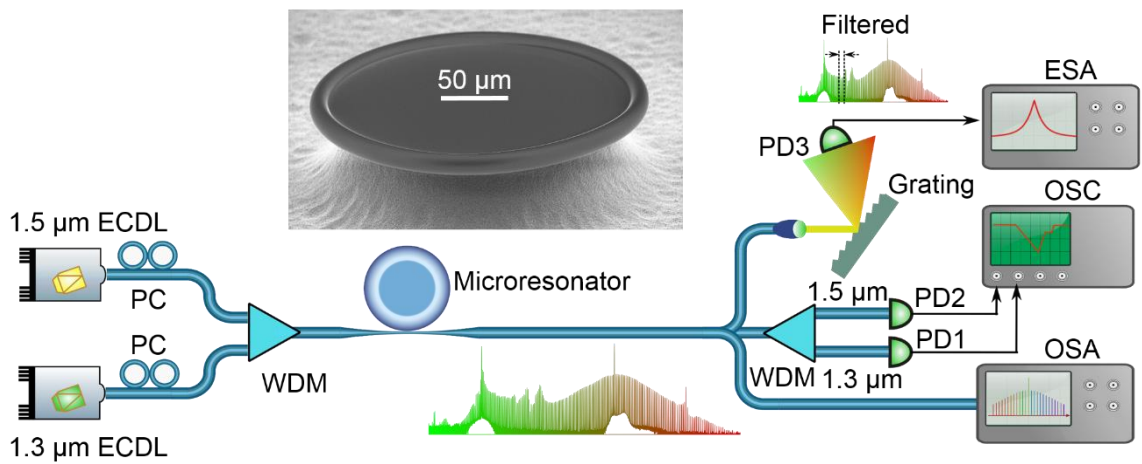

**Fig. 1: Experimental setup used for coherent, broadband microcomb generation via bichromatic pumping.** The 1.5-µm primary pump laser is used to generate a soliton microcomb, while the 1.3-µm auxiliary laser thermally stabilizes the resonator and at the same time generates a second frequency comb that broadens the spectrum. ECDL: external cavity diode laser; WDM: wavelength division multiplexer; PC: polarization controller; FBG: fiber Bragg grating; EDFA: Erbium-doped fiber amplifier; PD: photodetector; OSA: optical spectrum analyzer; OSC: oscilloscope; ESA: electronic spectrum analyzer. Inset: Scanning electron microscope image of one of the 250-µm-diameter microtoroids used in the experiments.

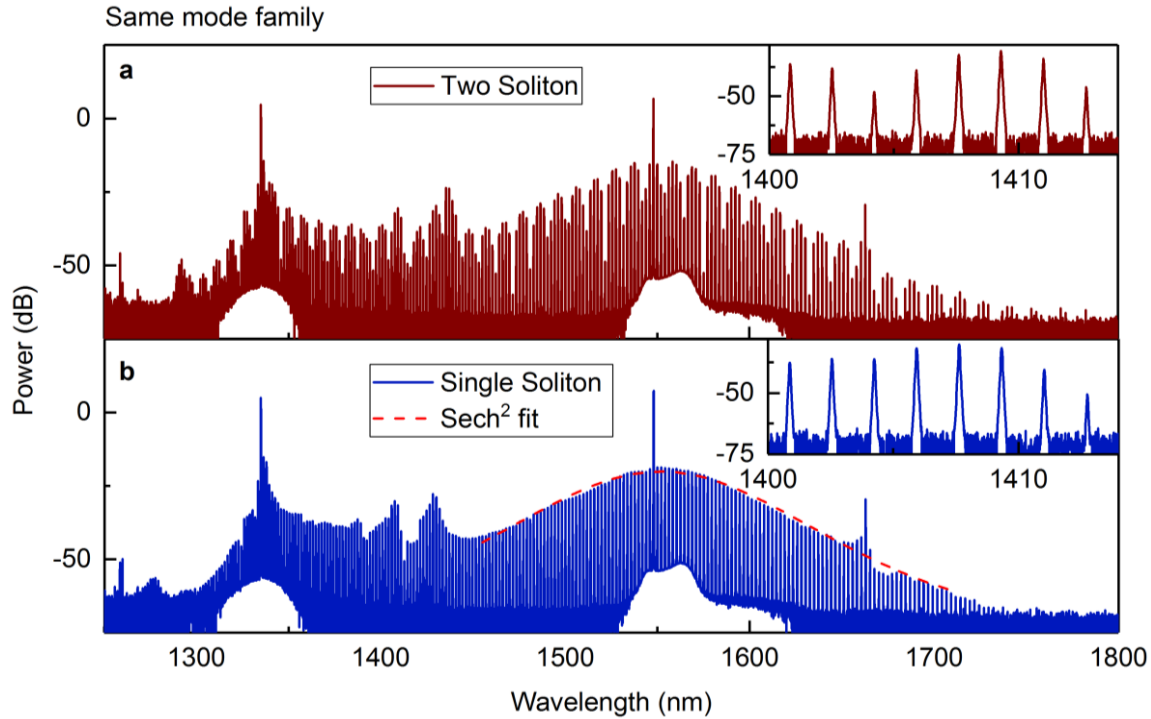

**Fig. 2: Optical spectra of coherently extended soliton states in a 250 micron microtoroid. a**, Optical spectrum of the two-soliton state. The primary pump power at 1550 nm is 250 mW and the auxiliary laser power at 1330 nm is 150 mW. Inset: zoomed-in spectrum in the spectral range 1400 nm to 1414 nm with a resolution of 0.02 nm. **b**, Optical spectrum of the single-soliton state. The red dashed line shows the predicted $sech^2$ envelope. Inset: zoomed-in spectrum between 1400 nm to 1414 nm with a resolution of 0.02 nm.

In the experiments, the frequency of the 1.3-μm auxiliary laser was first tuned downward in frequency into its resonance and then kept on the blue-detuned side of the resonance to passively stabilise the intracavity power and the temperature of the microresonator. The 1.5-μm primary pump laser was then tuned from the blue-detuned side of its resonance to the red-detuned soliton regime to generate a stable optical soliton. By optimizing the detuning and optical power of the 1.3-μm auxiliary laser, soliton states can be accessed by slowly tuning the primary pump laser frequency into the soliton regime[27]. In the experiments, ~250 mW of 1.5-μm primary pump light was used to generate the bright soliton microcomb, while ~150 mW of auxiliary pump light was used to passively compensate the thermally induced resonance shifts of the microresonator. Figure 2 shows the typical optical spectra of the microcomb in the two- and single-soliton states. Surprisingly, in the two-soliton state (Fig. 2a), we can see that the spectrum around the auxiliary pump frequency has the same pattern as the primary soliton spectrum, which suggests that the constituent pulses of the auxiliary comb waveform are locked to the primary solitons in the time domain. The inset of Fig. 2a shows the region of overlap (1400 nm to 1414 nm) of the optical spectrum in between the two pump lasers. Limited by the 0.02 nm resolution of the OSA, only a single peak was resolved for each pair of comb modes in the overlapping region. Figure

2b shows the spectrum of the microcomb in the single-soliton state. The optical spectrum around the primary pump wavelength has a smooth, sech$^2$-like envelope (red dashed line in Fig. 2b). The 3-dB bandwidth of the spectrum is approximately 7.4 THz, corresponding to a 50-fs optical pulse. Similarly, the inset of Fig. 2b shows the region of overlap of the single-soliton optical spectrum. Again, only one peak per FSR is observed by the OSA.

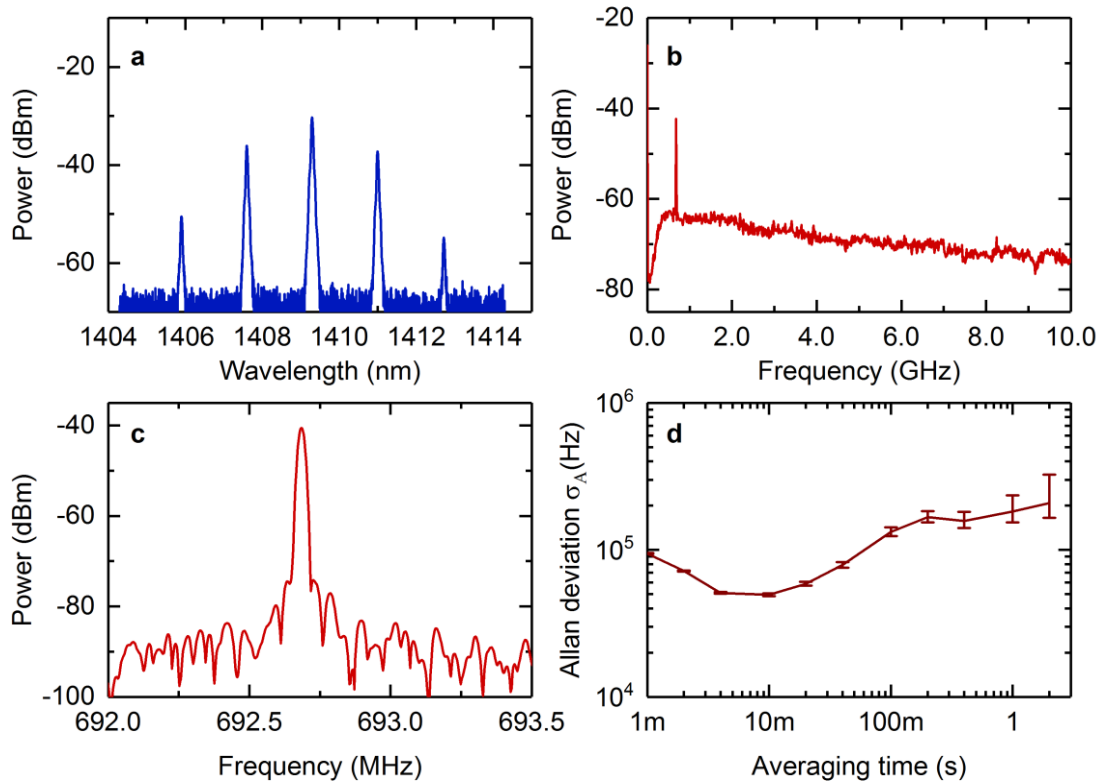

**Fig. 3: Optical spectrum in the overlapping region between the two frequency combs and corresponding RF spectra. a**, Filtered optical spectrum. **b**, RF spectrum of the beat note signal between the two frequency combs from DC to 10 GHz. **c**, RF spectrum of the beat signal with a 20 kHz resolution bandwidth. **d**, Allan deviation of the beat note signal.

To investigate whether the frequency comb spectra in the region of overlap shown in Fig. 2 are continuous or composed of two separate frequency combs, a grating (as shown in Fig. 1) was used to spectrally filter the overlapping portion of the spectrum. When operating the primary microcomb in the single-soliton state, the filtered spectrum (covering ~5 FSR) was sent to photodetector PD3 (see Fig. 3a). As shown in Fig. 3b, a single peak was observed in the radio frequency (RF) spectrum of the beat signal from DC to 10 GHz. Taking into account the 0.02 nm (4 GHz) resolution of the OSA, we can conclude that the optical spectra shown in Fig. 2 are composed of two different frequency combs resulting from the bichromatic pumping. Each comb line seen on the OSA in the overlapping region actually consists of two comb lines, one from each pump. Moreover, the existence of a single peak in the RF spectrum also reveals that the comb lines generated from the auxiliary laser have the exact same spacing as the primary soliton microcomb. Figure 3c shows the RF spectrum of the beat signal with a resolution bandwidth of 20 kHz. The beat note has a >35 dB signal-to-noise ratio, which implies the frequency comb generated from the auxiliary laser is low-noise and coherent. To characterise the relative drift between the primary soliton microcomb and the auxiliary comb, the beat signal was mixed down and measured with a frequency counter using a 1 ms gate time. Figure 3d shows the resulting Allan deviation of the beat note with fluctuations of ~180 kHz at 1 s. By locking this offset beat note to an RF reference, a low-noise extension of the microcomb can be generated.

By selecting a different auxiliary mode which does not belong to the soliton mode family, similar coherent extension of the microcomb optical spectrum results can be obtained, as shown in Fig. 4a,b. Figure 4a shows the optical spectrum of a multi-soliton state when the auxiliary laser is operating at 1334.6 nm. Similar to the previous results, the extended spectrum around the auxiliary pump frequency has the same pattern as that of the primary soliton, which indicates that the pulses at the auxiliary wavelength are locked into position with respect to the primary solitons, and hence that both frequency combs have the same repetition rate. Figure 4b shows the corresponding optical spectrum of the microcomb in the single-soliton state.

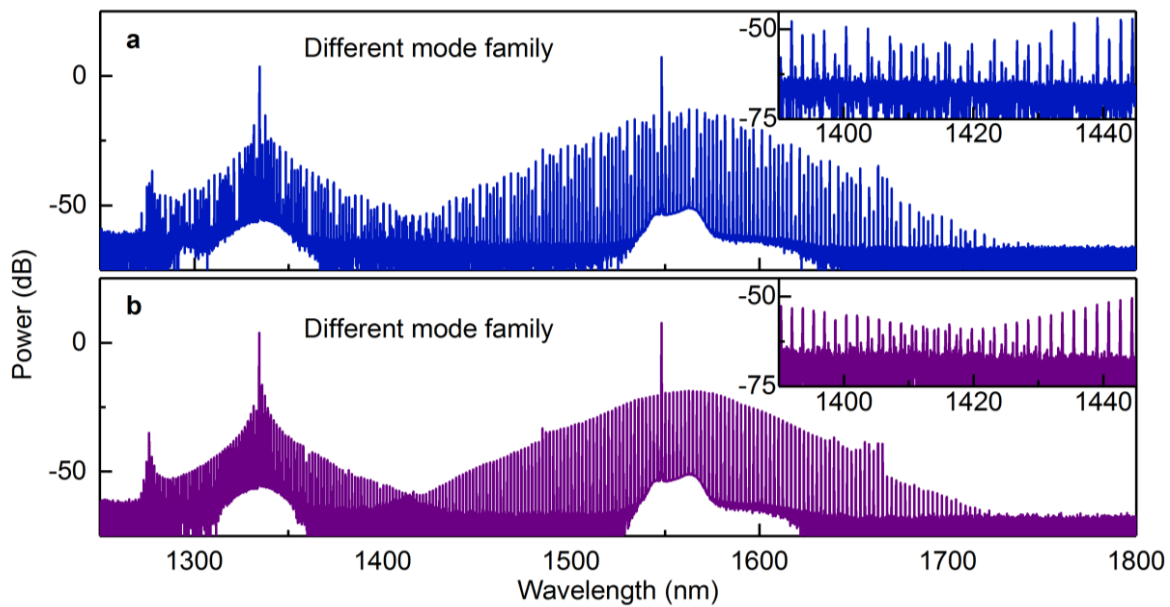

**Fig. 4: Spectral extension of microcombs across different mode families.** Panel **a** shows a multi-soliton state and panel **b** shows a measurement of a single soliton state. The envelope pattern is transferred between the two mode families. Insets show a zoom into the overlapping spectral region of the combs.

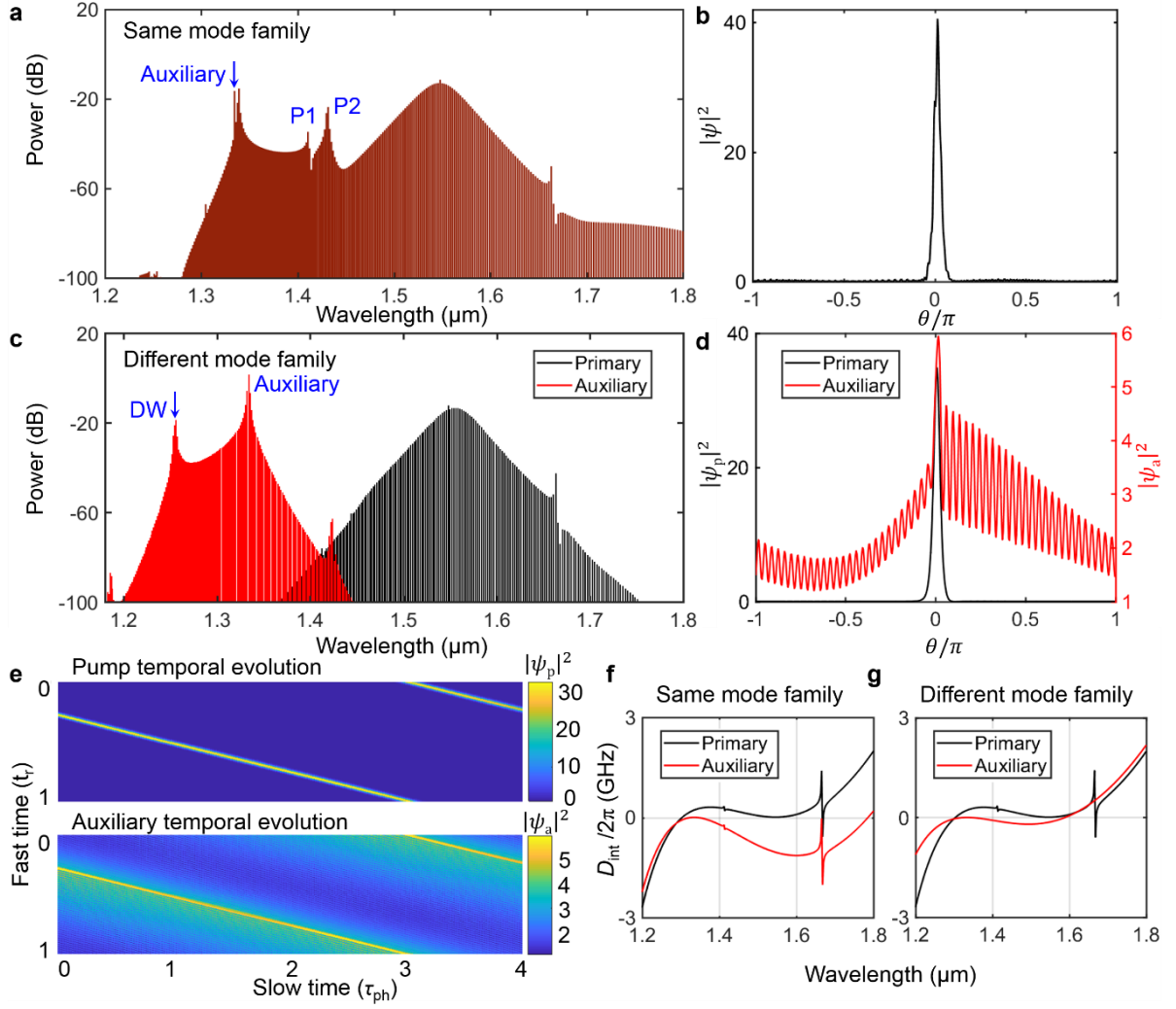

**Fig. 5: Numerical simulation of the temporal dynamics of intracavity optical fields. a**, Simulated intracavity optical spectrum of the microcomb in the single-soliton state with the auxiliary mode from the soliton mode family. **b**, Temporal waveform corresponding to the spectrum shown in **a**. **c**, Simulated intracavity optical spectrum of the microcomb in the single-soliton state with the auxiliary laser coupled into a different mode family from the soliton. DW: dispersive wave. **d**, Temporal waveforms (black, left axis) of the primary soliton and its induced bright auxiliary optical pulse (red, right axis) corresponding to the spectrum shown in **c**. **e**, Temporal waveform evolution of the primary soliton (upper panel) and bright auxiliary optical pulse (lower panel) for the same parameters as in **d**. **f, g**, Integrated dispersion $D_{\text{int}}$ of the 250-μm-diameter silica microtoroid relative to the primary (black) and auxiliary (red) pump modes, with the auxiliary mode from the same (**f**) and different (**g**) mode family, used for the numerical simulations.

In order to gain an insight into the temporal dynamics of the intracavity primary and auxiliary fields, we performed a number of numerical simulations based on the Lugiato-Lefever equation (LLE)[29-31], for the cases in which the auxiliary laser is coupled to an optical mode either belonging or not belonging to the soliton mode family. Figure 5a represents the simulated intracavity optical spectrum in the first case when the auxiliary mode is from the soliton mode family and the primary soliton is in a single-soliton state. The simulated results are in good agreement with the experimental measurements (Fig. 2b), with two distinct peaks (dispersive waves, DWs) in the overlapping region of the optical spectrum[32]. The simulated results are explained in more detail in the "Methods" section and in the Supplemental Information. From the simulations, we determine that the left peak, marked as P1 in Fig. 5a, is induced by an avoided mode crossing[33], while the right one, marked as P2, is generated by the frequency-dependent dispersion of the resonator modes, where phase matching between the auxiliary optical pulse and the DW occurs[34,35]. Once stable soliton states are accessed, the repetition rate of the auxiliary optical pulse is coerced to synchronise with the primary soliton, changing the effective local D1 coefficient for the auxiliary frequency comb and thereby the phase matching for the P2 DW. Even though the integrated dispersion[32] of the auxiliary modes does not cross zero around 1430 nm (see Methods), the phase matching condition for the P2 DW still occurs due to this alteration of the repetition rate. Figure 5b is the simulated temporal waveform of a microcomb in the single-soliton state. It should be noted that the spatiotemporal modulations in the pulse pedestal are induced by the beat signals between the primary pump and the strongest auxiliary sideband, as well as the two DWs.

Figure 5c shows the simulated intracavity optical spectrum when the primary soliton is in a single-soliton state in the case where the auxiliary laser is coupled to a mode that does not belong to the soliton mode family. These simulated results are also in good agreement with the measurements shown in Fig. 4b. Figure 5d depicts the simulated temporal waveform of a single soliton (black, left axis) generated by the primary pump laser. Surprisingly, a bright optical pulse (red, right axis) is generated in the normal dispersion regime by the auxiliary laser. It should be noted that the bright optical pulse is formed and trapped by the bright soliton via XPM. This indicates that the auxiliary frequency comb has the same repetition rate as the primary soliton. Similar to the experimentally obtained optical spectrum in Fig. 4b, there is also a peak (DW, marked with arrow) in the simulated auxiliary optical spectrum, which induces the fast oscillations in the pedestal of the bright auxiliary optical pulse as shown in Fig. 5d. Figure 5e shows the corresponding evolution of the temporal waveforms. We can see that the bright auxiliary optical pulse propagates with the same group velocity as the primary soliton, which verifies the locking of the repetition rate.

It should be noted that, as shown in Fig. 5c, when using an auxiliary mode that does not belong to the soliton mode family, the centre frequency of the primary soliton comb envelope slightly shifts to the red side due to XPM with the auxiliary frequency comb, which is consistent with the experimental optical spectra shown in Fig. 4a and b.

## Discussion

We have demonstrated that the optical spectrum of a soliton microcomb can be coherently extended through bichromatic pumping. The first pump laser generates a soliton microcomb in the anomalous dispersion regime, while the second laser both thermally stabilizes the microresonator and produces a second frequency comb at a repetition rate that is passively synchronised with the main soliton through cross-phase modulation. Our numerical simulations of the spectrally extended microcombs are in excellent agreement with the experimental results. Further simulations reveal the shape of the optical pulses that are generated in the normal dispersion regime via bichromatic pumping. These findings are further confirmed even when using auxiliary optical modes with different polarisation to the primary comb. By measuring or locking the offset frequency between the two combs, this technique can be used to generate a coherently extended frequency comb without requiring advanced dispersion engineering of the microresonators. In particular this technique might allow bridging of spectral regions with unfavourable dispersion for broadband frequency comb generation. Moreover, the demonstrated method can be used to enhance the power of comb modes in selected spectral regions, which has applications in optical clocks and optical spectroscopy. By locking both pump frequencies to atomic frequency references, the repetition rate of the microcomb is directly related to the difference between the two pumping frequencies, and can be used to create an optical clock. We believe that the observed synchronisation between microcombs in different spectral regions in a microresonator will be of wide interest for optical frequency metrology and spectroscopy using microresonator-based frequency combs.

## Methods

**Numerical Simulations.** Figures 5f and g show the integrated dispersion[32] $D_{\text{int}}$ profiles relative to the primary (black curve) and auxiliary (red curve) pump modes for the 250-µm-diameter silica microtoroid. In Fig. 5f, the auxiliary laser is coupled into the same mode family and in Fig 5g into a different mode family compared to the primary pump laser (see the Supplemental information for further details). The combined material and geometrical dispersion is given as:

$$D_{\text{int},*}(\mu) = \omega_{*,\mu} - (\omega_{*,0} + D_{*,1}\mu) = \frac{D_{*,2}}{2!}\mu^2 + \frac{D_{*,3}}{3!}\mu^3, \tag{1}$$

where $\mu$ is the mode number relative to either the primary or auxiliary pump mode as relevant (for which $\mu = 0$), $\omega_{*,\mu}$ is the resonance frequency of the that mode, '*' refers to either the primary ('p') or auxiliary ('a') comb, $D_{*,1}/2\pi$ is the FSR of the resonator at either the primary ($D_{p,1}/2\pi$) or auxiliary ($D_{a,1}/2\pi$) pump mode, and $D_{*,2}$, $D_{*,3}$ are coefficients of second- and third-order dispersion, respectively. From Fig. 5f and g, $D_{\text{int},p}(\mu)$ (black curve) is approximately parabolic close to the primary pump mode, indicating anomalous group velocity dispersion ($D_{p,2} > 0$), which supports the formation of bright solitons. Due to the third-order dispersion, $D_{\text{int},p}(\mu)$ crosses zero again at around 1300 nm, which can lead to the appearance of a DW[32]. In order to accurately simulate the experimental results shown in Fig. 2 and 4, we added avoided mode crossings around 1410 nm and 1663 nm[33]. In contrast, $D_{\text{int},a}(\mu)$ (red curve in Fig. 5f and g) shows normal group velocity dispersion around the auxiliary mode.

In the case where the auxiliary laser is coupled into an optical mode from the soliton mode family, we solved the simulations by using a single generalized LLE[22]:

$$\frac{\partial \psi(\theta,\tau)}{\partial \tau} = -(1+i\alpha_p)\psi + i|\psi|^2\psi - \sum_{n=2}^{N\geq 2}(-i)^{n+1}\frac{\beta_{p,n}}{n!}\frac{\partial^n \psi}{\partial \theta^n} + F(\theta,\tau), \quad (2)$$

$$F(\theta,\tau) = F_p + F_a \exp\left(i\mu_a\theta - i(\alpha_a - \alpha_p)\tau - i\frac{2D_{int,p}(\mu_a)}{\Delta\omega_0}\tau\right), \quad (3)$$

where $\tau$ is the slow time, normalized to twice the photon lifetime ($\tau_{ph}$), $\theta$ is the azimuthal angle in a frame co-rotating at the average of the primary and auxiliary group velocities. $F_p$, $F_a$ are the dimensionless external pump amplitudes, and $\psi(\theta,\tau)$ is the intracavity field envelope driven by both pump terms $F_p$, $F_a$. $\alpha_p$ and $\alpha_a$ are the frequency detunings from the primary and auxiliary pump lasers with respect to their respective resonance frequencies and both normalized to half of the full-width at half-maximum (FWHM) of the resonance ($\Delta\omega_0$). $\beta_{p,n}$ are the $n^{th}$-order dimensionless dispersion coefficients at the primary pump mode, normalized as $\beta_{p,n} = -2D_{p,n}/\Delta\omega_0$. $\mu_a$ is the mode number of the auxiliary mode relative to the primary pump mode with the value $\mu_a = 120$. The LLE simulations are performed numerically using the split-step Fourier method. As in the experiments, the primary pump frequency is scanned from the blue- to the red-detuned side of its resonance until stable soliton states are generated. During frequency scanning of the primary pump laser, the auxiliary laser detuning ($\alpha_a$) is kept constant. In the simulations, the parameters are $\beta_{p,2} = -0.2140$, $\beta_{p,3} = 0.0046$, $|F_a|^2 = 8$, $|F_p|^2 = 16$. Calculated from these dispersion coefficients, the group velocity mismatch $\gamma$ between the primary and auxiliary pump modes is -3.62, normalized to the FWHM as $\gamma = (D_{a,1} - D_{p,1})/\Delta\omega_0$, meaning that $D_{p,1} > D_{a,1}$. The negative sign of $\gamma$ causes the P2 DW to appear on the red side of the auxiliary mode (see the Supplemental information for details).

For the case in which the auxiliary laser is coupled into a mode that does not belong to the soliton mode family, we solve a system of two simultaneous generalized LLEs, which are expanded with XPM and a group velocity mismatch $\gamma$ between the primary and auxiliary pump modes [25,26]:

$$\frac{\partial \psi_p(\theta,\tau)}{\partial \tau} = -(1+i\alpha_p)\psi_p + i\left(|\psi_p|^2 + \sigma|\psi_a|^2\right)\psi_p - \sum_{n=2}^{N\geq 2}(-i)^{n+1}\frac{\beta_{p,n}}{n!}\frac{\partial^n \psi_p}{\partial \theta^n} + \gamma\frac{\partial \psi_p}{\partial \theta} + F_p, \quad (4)$$

$$\frac{\partial \psi_a(\theta,\tau)}{\partial \tau} = -(1+i\alpha_a)\psi_a + i\left(|\psi_a|^2 + \sigma|\psi_p|^2\right)\psi_a - \sum_{n=2}^{N\geq 2}(-i)^{n+1}\frac{\beta_{a,n}}{n!}\frac{\partial^n \psi_a}{\partial \theta^n} - \gamma\frac{\partial \psi_a}{\partial \theta} + F_a, \quad (5)$$

where $\psi_p(\theta,\tau)$, $\psi_a(\theta,\tau)$ are the intracavity primary and auxiliary field envelopes, respectively. $\sigma$ is the XPM coefficient (2/3 for orthogonal polarisation and 2 for the same polarisation assuming perfect spatial mode overlap, otherwise less). Although $\sigma$ is difficult to determine experimentally, varying it from 2/3 to 2 does not affect the simulation much, resulting mostly in slightly different amounts of redshift of the centre frequency of the soliton comb envelope. $\beta_{a,n}$ are the $n^{th}$-order dimensionless dispersion coefficients, normalized as $\beta_{a,n} = -2D_{a,n}/\Delta\omega_0$. In the simulations, the parameters are $\beta_{p,2} = -0.2140$, $\beta_{p,3} = 0.0046$, $\beta_{a,2} = 0.1441$, $\beta_{a,3} = 0.0031$, $|F_a|^2 = 8$, $|F_p|^2 = 16$, $\sigma = 1.5$ and $\gamma = 2.5$. Here, the positive sign of $\gamma$ determines the appearance of the DW on the blue side of the auxiliary mode as shown in Fig. 5c (see the Supplemental information for details).

## Acknowledgements

This work is supported by H2020 Marie Sklodowska-Curie Actions (MSCA) (748519, CoLiDR), a Horizon 2020 Marie Sklodowska-Curie grant (GA-2015-713694), the National Physical Laboratory Strategic Research Programme, a H2020 European Research Council (ERC) grant (756966, CounterLight), and the Engineering and Physical Sciences Research Council (EPSRC). J.M.S. acknowledges funding via a Royal Society of Engineering fellowship.

## Supplementary Information

### 1. Optical spectrum of chaotic states

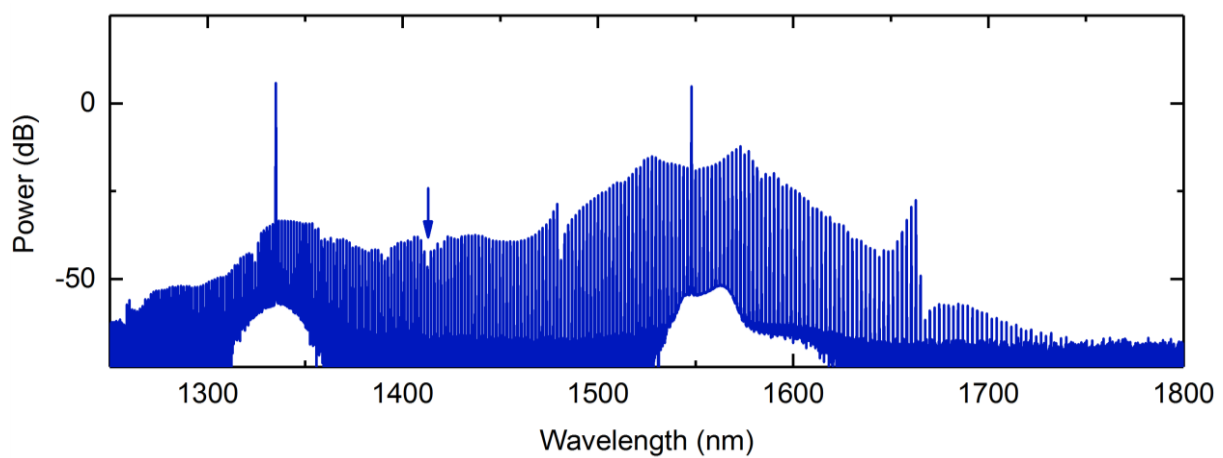

**Fig. S1: Experimental optical spectrum of the chaotic states with the auxiliary laser pumping an optical mode (~1335 nm) from the soliton mode family.** These optical spectra show an avoided mode crossing around 1410 nm but no dispersive wave around 1430 nm.

Figure S1 shows the optical spectrum of the chaotic states with the auxiliary laser pumping an optical mode (~1335 nm) from the soliton mode family. The optical spectrum shows a very strong avoided mode crossing (AMC) around 1663 nm and a weak one around 1410 nm (marked with arrow). However, there is no dispersive wave around 1430 nm, which is different from the soliton spectra shown in Fig. 2 of the main text. Therefore in the simulation, as shown in Fig. S2 here and Fig. 5f in the main text, only one AMC is added around 1410 nm (relative mode number 71 to the primary pump mode) without a phase-matching point for the DW around 1430 nm in the integrated dispersion $D_{\text{int}}$.

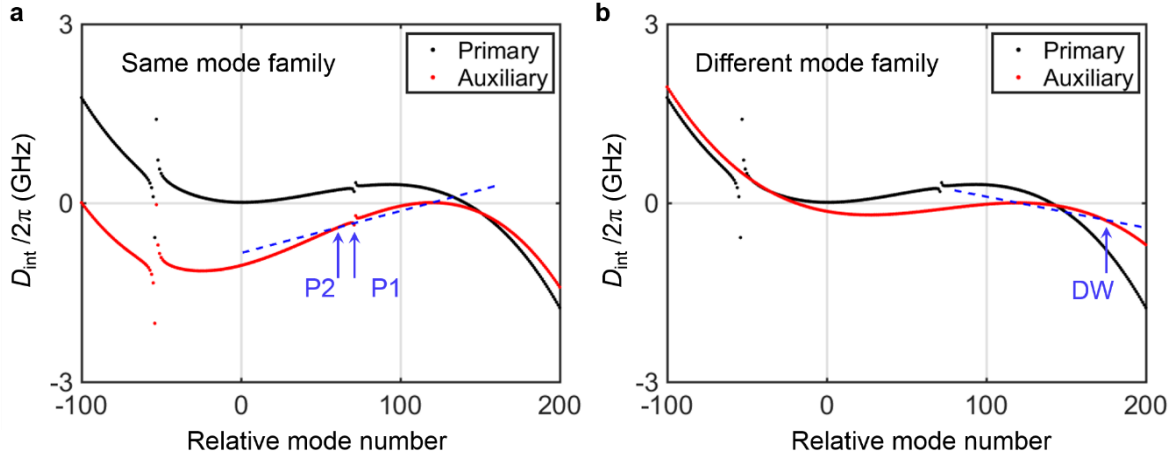

**Fig. S2: The integrated dispersion $D_{int}$ ($\mu$) of the 250-μm-diameter silica microtoroid relative to the primary (blue curve) and auxiliary (red curve) pump modes with the auxiliary laser pumping a mode from (a) the same and (b) a different mode family to the soliton, plotted against the relative mode number.** The primary pump mode number is 0 while the auxiliary mode is chosen as 120. The dashed lines in **a**, **b** show the phase-matching points between the auxiliary bright optical pulse and auxiliary dispersive waves.

## 2. Phase matching between the auxiliary bright optical pulse and the auxiliary dispersive wave

For the auxiliary frequency comb, a dispersive wave is expected to occur at a mode number $\mu'$ for which the comb mode is exactly on resonance with the resonator mode. Thus $\mu'$ satisfies the condition

$$\omega_a + \omega_{rep}\mu' = \omega_{a,0} + D_{a,1}\mu' + D_{int,a}(\mu'). \tag{S1}$$

Here, $\omega_a$ is the auxiliary laser frequency and $\omega_{rep}$ is the repetition rate of the auxiliary frequency comb. $\omega_{a,0}$ is the auxiliary mode frequency, $D_{a,1}/2\pi$ is the FSR of the resonator at the auxiliary mode, and $D_{int,a}(\mu')$ is the integrated dispersion relative to the auxiliary pump mode, defined in the main text. Eq. (S1) can be rearranged to give

$$D_{int,a}(\mu') = (\omega_{rep} - D_{a,1})\mu' + \omega_a - \omega_{a,0}, \tag{S2}$$

which is solved graphically in Fig. S2.

If the repetition rate of the auxiliary frequency comb equals to the $D_{a,1}$, a dispersive wave will occur at a mode number of around -100, as shown in Fig. S2a for the case in which the auxiliary laser is coupled into an optical mode from the soliton mode family, and at the mode number of around -20, as shown in Fig. S2b for the case in which the auxiliary laser is coupled into a mode different from the soliton mode family. However, due to XPM with the primary pump, the repetition rate of the auxiliary frequency comb is synchronized with that of the primary soliton, which mainly depends on $D_{p,1}$. Therefore, in the case in which the auxiliary laser pumps a mode belonging to the soliton mode family, the slope of the r.h.s of Eq. S2 is positive, as $D_{p,1}$ is larger than $D_{a,1}$. This can be interpreted by the dashed line plotted in Fig. S2a, inducing phase-matching at a mode number of 61 (marked with an arrow), corresponding to the DW marked P2 in Fig.5a of the main text. Similarly, in the case in which the auxiliary laser pumps a mode that is not from the soliton mode family, the slope of the r.h.s of Eq. S2 is negative due to $D_{p,1}$ being smaller than $D_{a,1}$. This induces a phase-matching point at the mode number of 174, as shown in Fig. S2b, corresponding to the DW in Fig. 5c of the main text.

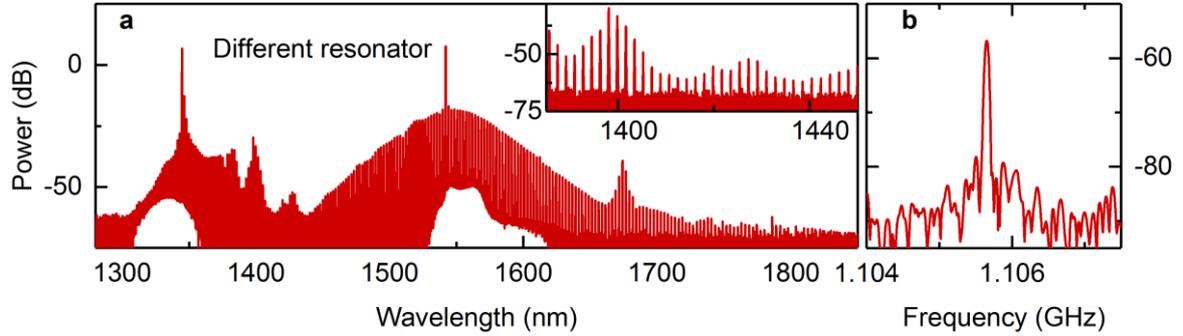

**Fig. S3: Spectral extension and synchronization of microcombs in a different microtoroid. a**, Optical spectrum of a single-soliton state generated using a different microtoroid with the primary and auxiliary pumps coupled into the same optical mode family. **b**, RF spectrum of the beat signal between the primary soliton microcomb and the auxiliary frequency comb with a 50 kHz resolution bandwidth, when the primary soliton microcomb is in a multi-soliton state.

### 3. Spectral extension and synchronization of microcombs in a different microtoroid

To further confirm our findings, we performed similar experiments using a different microtoroid. Selecting an optical mode from the soliton mode family for the auxiliary pump, the extended optical spectrum of a single-soliton state microcomb is shown in Fig. S3a. The inset is the optical spectrum in the region of overlap. In the same way as with Fig. 2 of the main text, only a single set of frequency comb modes is observed due to the limited resolution of the OSA. To detect the offset signal between the primary soliton microcomb and the auxiliary frequency comb, the primary soliton microcomb was operated in a multi-soliton state to increase the optical power in the region of overlap. Figure S3b is the corresponding RF spectrum of the offset signal. Indeed, a single RF peak at around 1.1 GHz is observed, further confirming that the repetition rates of the two combs are locked to one another and that this method could be used for the coherent extension of microcombs.